\documentclass[aps,prl,twocolumn,10pt,superscriptaddress]{revtex4}

\usepackage[paperwidth=210mm,paperheight=297mm,centering,hmargin=2cm,vmargin=2.1cm]{geometry}
\usepackage[utf8]{inputenc}
\usepackage[english]{babel}
\usepackage{amsmath}
\usepackage{mathtools}
\usepackage{enumitem}
\usepackage{amssymb,amsfonts}

\usepackage[perpage,symbol*]{footmisc}
\usepackage[final]{graphicx}
\usepackage{pstricks}

\usepackage{txfonts}
\usepackage{cancel}
\usepackage[colorlinks=true,linkcolor=magenta,urlcolor=black,hyperindex]{hyperref}

\begin{document}

%% -------------------Alternative commands-------------------------

\newcommand{\ket}[1]{\vert#1\rangle}
\newcommand{\tket}[1]{\vert{\boldsymbol#1}\rangle}
\newcommand{\tbra}[1]{\langle{\boldsymbol#1}\vert}
\newcommand{\bra}[1]{\langle#1\vert}
\newcommand{\braket}[2]{\langle#1\vert#2\rangle}
\newcommand{\ketbra}[2]{\vert#1\rangle\langle#2\vert}
\newcommand{\cub}[1]{\left(#1\right)}
\newcommand{\sqb}[1]{\left[#1\right]}
\newcommand{\pli}[1]{\hat{\sigma}_{#1}}
\newcommand{\rvec}[1]{\overset{\shortleftarrow}{#1}}
\newcommand{\uv}[1]{\vec{x}_{(#1)}} % for unit vector
\newcommand{\dv}[1]{\underset{\vec{}\hspace{0.025cm}}{x}^{(#1)}} % Dual Vector
\newcommand{\uve}[1]{\vec{e}_{(#1)}} % for unit vector
\newcommand{\dve}[1]{\underset{\vec{}\hspace{0.025cm}}{e}^{(#1)}} % Dual Vector
\newcommand{\ddve}[1]{\underset{\vec{}\hspace{0.025cm}}{\dot{e}}^{(#1)}} % Dual Vector
\newcommand{\aff}[3]{\Gamma^{#1}_{\;\;#2#3}}
\newcommand{\pd}[2]{\frac{\partial #1}{\partial #2}} % for partial derivatives
\newcommand{\quat}[1]{\hat{\sigma}_{(#1)}}

%% -------------------Alternative commands-------------------------

\title{Who's afraid of the Unit Quaternion ?}

\author{Brian O'Sullivan}
% \email{brianosullivan@meditations-on-geometry.net}
\affiliation{25 Copperhill, Ballintemple, Cork, Ireland.}

% \date{\today}

\begin{abstract}
 Far from being just a 2-level Quantum system the Qubit is a Unit Quaternion, also known as a Spinor. 
 Therefore it follows that the Qubit is a 4-dimensional vector which traces a path on the surface of the unit 3-sphere. 
 This is the meaning of the global phase.
\end{abstract}

\maketitle

\noindent\underline{Proposition:} The Qubit $\ket{\Psi^\pm}\in\mathbb{C}^2$ is a Unit Quaternion.

\noindent\underline{Proof:} The Qubit is a 2-component column or row vector with complex numbers $\alpha$ and $\beta$ satisfying $|\alpha|^2+|\beta|^2=1.$
There are two orthonormal representation of the Qubit, respectively denoted by the {\it kets} $\ket{\Psi^+}=\binom{\alpha}{\beta}$ and $\ket{\Psi^-}=\binom{-\beta^*}{\alpha^*}$.
The {\it bras} $\bra{\Psi^\pm}$ are the transpose conjugate of the kets, i.e. $\bra{\Psi^\pm}=\cub{\ket{\Psi^\pm}}^\dagger$.
The Qubit satisfies the orthonormality relations $\braket{\Psi^\pm}{\Psi^\pm}=1$ and $\braket{\Psi^\pm}{\Psi^\mp}=0.$ 

The Qubit is parameterised in terms of the 3 angles of the 3-sphere as,
\begin{equation}
\label{eq:qubit}
 \ket{\Psi^+(t)}\;=\;
 \begin{pmatrix}
  e^{-\imath\tfrac{\omega+\phi}{2}}\cos\cub{\tfrac{\theta}{2}} \\
  e^{-\imath\tfrac{\omega-\phi}{2}}\sin\cub{\tfrac{\theta}{2}}
 \end{pmatrix};\;\;
 \ket{\Psi^-(t)}\;=\;
 \begin{pmatrix}
  -e^{\imath\tfrac{\omega-\phi}{2}}\sin\cub{\tfrac{\theta}{2}} \\
  e^{\imath\tfrac{\omega+\phi}{2}}\cos\cub{\tfrac{\theta}{2}}
 \end{pmatrix}.
\end{equation}
$\{\omega(t),\phi(t),\theta(t)\}$ are the global phase, the azimuthal angle and the polar angle.

The Unit Quaternion is a 2x2 complex matrix which takes the form,
\begin{equation}
\label{eq:quaternion}
 \hat{U}(t)\;=\;
 \begin{pmatrix}
  a+\imath b & c+\imath d \\
  -c+\imath d & a - \imath b
 \end{pmatrix}.
\end{equation}
Where $\{a(t),b(t),c(t),d(t)\}\in\mathbb{R}$, and $\hat{U}(t)\in\mathbb{R}^4.$ The Unit Quaternion %has unit determinant and 
is normalised with respect to the Hermitian inner product $\hat{U}\hat{U}^\dagger=\quat{1},$ where $\quat{1}$ is the 2x2 identity matrix.

Consider that the Qubit can be written in Quaternion form as 
\begin{align}\nonumber 
\hat{\Psi}(t)\;&=\;\cub{\ket{\Psi^+}\;\;\ket{\Psi^-}},\\
\label{eq:spinor}
 \hat{\Psi}(t)\;&=\;
 \begin{pmatrix}
  e^{-\imath\tfrac{\omega+\phi}{2}}\cos\cub{\tfrac{\theta}{2}} & -e^{\imath\tfrac{\omega-\phi}{2}}\sin\cub{\tfrac{\theta}{2}} \\
  e^{-\imath\tfrac{\omega-\phi}{2}}\sin\cub{\tfrac{\theta}{2}} & e^{\imath\tfrac{\omega+\phi}{2}}\cos\cub{\tfrac{\theta}{2}}
 \end{pmatrix}.
\end{align}
% In this parameterized form the Unit Quaternion is known as a Spinor. 
The difference between the Dirac {\it bra-ket} form \eqref{eq:qubit}, 
and the Quaternion form \eqref{eq:spinor} of the Qubit is simply a matter of notation.
Given that both representations of the Qubit \eqref{eq:qubit} and \eqref{eq:spinor} 
evolve from their initial state via the Unitary matrix, 
$$\ket{\Psi^\pm(t)}\;=\;\hat{U}(t)\ket{\Psi^\pm(0)};\qquad\hat{\Psi}(t)\;=\;\hat{U}(t)\hat{\Psi}(0),$$
and satisfy Schr\"odinger's equation,
$$\imath\ket{\dot{\Psi}^\pm(t)}\;=\;\mathcal{\hat{H}}(t)\ket{\Psi^\pm(t)};\qquad
\imath\dot{\hat{\Psi}}(t)\;=\;\mathcal{\hat{H}}(t)\hat{\Psi}(t),
$$
$\mathcal{\hat{H}}(t)=\imath\dot{\hat{U}}\hat{U}^\dagger,$ then it is clear 
the Dirac \eqref{eq:qubit} and Quaternion \eqref{eq:spinor} representations of the Qubit are equivalent. Q.E.D. 
% Quot erat demonstrandum.

Unit Quaternions \eqref{eq:quaternion} with initial value equal to the identity matrix $\hat{U}(0)=\quat{1},$ are known as the Unitary matrix, 
whereas Unit Quaternions of the form \eqref{eq:spinor} with an initial orientation not along the poles, i.e.
$\theta(0)\neq 0$, etc.,
% for $n\in\mathbb{N}$, 
are known as Spinors.

\noindent\underline{Conclusion:} The Qubit $\ket{\Psi^\pm}\in\mathbb{C}^2$ is a Unit Quaternion.

\noindent\underline{Discussion:} Under the $\mathbb{S}^3\xmapsto{\mathbb{S}^1}\mathbb{S}^2$ Hopf-Fibration, 
defined by $$\mathcal{\hat{R}}(\theta,\phi)\;=\;\hat{\Psi}(t)\frac{\quat{z}}{2}\hat{\Psi}^\dagger(t),$$
the global phase is identified as a natural hidden variable \cite{wharton:15}.
The Spinor $\hat{\Psi}$ describes a path in $\mathbb{S}^3$ and the Bloch vector $\mathcal{\hat{R}}$ describes a path in $\mathbb{S}^2,$  
parameterised by the polar and azimuthal angles.
$\mathbb{S}^1$ is the unit circle $e^{\imath\tfrac{\omega}{2}}$, parameterised by the global phase. 
This is a fiber bundle consisting of the
intrinsic parameters 
which are 
the global, geometric and dynamic phases \cite{Aharonov:87,osull:15b} linking 
% the bases spaces 
$\mathbb{S}^3$ and $\mathbb{S}^2$.

In \cite{osull:16} the global phase was explored as a possible explanation for the nature of the intrinsic spin of the fundamental particles. 
It was shown that the global phase of all closed paths in $\mathbb{S}^2$ is quantized by $2n\pi$ for $n\in\mathbb{N}$, 
thus rendering the $\mathbb{S}^1$ fibration $e^{\imath\tfrac{\omega}{2}}=\pm1.$ 
This allows the closed $\mathbb{S}^2$ paths to be classified as either bosonic or fermionic according to whether $e^{\imath\tfrac{\omega}{2}}=+1,$ 
or $e^{\imath\tfrac{\omega}{2}}=-1.$ 
We can however consider an equivalent Hopf-Fibration $\mathbb{S}^3\xmapsto{\mathbb{S}^1}\mathbb{S}^2$ defined by
$$\mathcal{\hat{Q}}(\theta,\omega)\;=\;\hat{\Psi}^\dagger(t)\frac{\quat{z}}{2}\hat{\Psi}(t).$$
In this case the Bloch vector is parameterised by the polar and global phases whereas the $\mathbb{S}^1$ fibration is parameterised by the azimuthal angle.
Here the global and azimuthal angles have reversed their roles as the azimuthal angle is now the intrinsic parameter. 
Therefore the global and azimuthal angles have equivalent roles, 
since the Hopf-Fibration offers two different but equally valuable perspectives of the same Spinor in $\mathbb{R}^3$.

The polar angle $\theta$ is defined with respect to the axis 
which penetrates the poles of $\mathbb{S}^3$ and $\mathbb{S}^2$. 
The poles of the 3-sphere and 2-sphere are coordinate singularities, e.g. 
when $\theta=0$, the global and polar angles are not defined.
The study of the 2-Level Quantum system is the study of the polar angle $\theta$, which describes the Rabi oscillations of the Qubit.
This angle describes the longitudinal mode of the Spinor, whereas the global $\omega$ and azimuthal $\phi$ angles describe the transverse modes. 
The transverse modes are the Hidden Variables of Quantum Mechanics, about which relatively little is known. 
The Hopf-Fibration is a projection from 
4-dimensions to 3-dimensions
% $\mathbb{R}^4\to\mathbb{R}^3$ 
which {\it `rolles up'} either one of these transverse 
modes - as a Fiber bundle - which is then encoded in the $\mathbb{S}^2$ path through the geometric and dynamic phases \cite{Aharonov:87,osull:15b}.  
Theories concerning the intrinsic spin 
of the fundamental particles
derived from the Unit Quaternion should regard both parameters of the transverse modes on an equal footing \cite{osull:16}. 

\section*{Acknowledgments}

This article is dedicated to all those Quantum mathematicians who, over the years,
I have begged and pleaded with to look at my calculations - 
to which they would half-heartedly humour me and then invariably refuse. 
I eventually came to realise that this rejection is not related to the quality of my work, 
rather it is rooted in fear. 
The petrifying fear that if they were to {\it look} and {\it measure} the 
accuracy of my calculations, then the quantum superposition would inevitably {\it collapse}, 
taking with it a massive body of literature and who knows ... maybe even the Quantum Theory itself.

%% ------------------------------------------------------------------------------------------------ 

\vspace*{-6pt}
\centerline{\rule{72pt}{0.4pt}}


\begin{thebibliography}{99}
  
\bibitem{wharton:15} K. B. Wharton and D. Koch, {\it ``Unit Quaternions and the Bloch Sphere''}
Journal of Physics A: Math. Theor. {\bf48} 235302 (2015).

\bibitem{Aharonov:87} Y. Aharonov and J. Anandan, {\it ``Phase Change During A Cyclic Quantum Evolution''}
Physical Review Letters {\bf 58} 1593 (1987).

\bibitem{osull:15b} Brian~O`Sullivan, {\it ``Vectors, Spinors and Galilean Frames''}
Meditations On Geometry {\bf01} 012028 (2015).

\bibitem{osull:16} Brian~O`Sullivan, {\it ``The Hopf-Fibration and Hidden Variables in Quantum and Classical Mechanics''}
Meditations On Geometry {\it 1$^{st}$ Meditation} (2016).

\end{thebibliography}
\end{document}